\documentclass[12pt]{article}
\usepackage{epsf,epsfig}
\usepackage{amsmath,amsfonts,amssymb}
\usepackage{cite}
\usepackage{breqn}

\usepackage[english]{babel}
\usepackage{hyperref}
\usepackage{graphicx,color}
\definecolor{link}{rgb}{.1,.1,.95}
\hypersetup{colorlinks=true,linkcolor=link,citecolor=link,urlcolor=link,linktocpage}

\newcommand{\be}{\begin{equation}}
\newcommand{\ee}{\end{equation}}
\newcommand{\ben}{\begin{eqnarray}\displaystyle}
\newcommand{\een}{\end{eqnarray}}

\newcommand{\bea}[1]{\begin{eqnarray}\label{#1} }
\newcommand{\eea}{\end{eqnarray}}

\newcommand{\refb}[1]{(\ref{#1})}

\newcommand{\sectiono}[1]{\section{#1}\setcounter{equation}{0}}

\newcommand{\del}{\partial}

\def\boxempty{{\,\lower0.9pt\vbox{\hrule \hbox{\vrule height 0.25 cm
\hskip 0.25 cm \vrule height 0.25 cm}\hrule}\,}}

\def\one{{\hbox{ 1\kern-.8mm l}}}
\def\zero{{\hbox{ 0\kern-1.5mm 0}}}

\begin{document}
\begin{titlepage}
\thispagestyle{empty}

\title{
{\Large\bf Stability of the Travelling Front of a Decaying Brane}
}

\author{
{\large\bf Debashis Ghoshal
and
Preeda Patcharamaneepakorn}\\
{}\\
{\large\it School of Physical Sciences}\\
{\large\it Jawaharlal Nehru University}\\
{\large\it New Delhi 110067, India}\\
{}\\
{\tt dghoshal@mail.jnu.ac.in}\\
{\tt preeda.pat@gmail.com}\\
{}\\
}

\bigskip\bigskip

\date{%
%
\bigskip\bigskip\bigskip\bigskip
\begin{quote}
\centerline{\large\bf Abstract:}
{\small
The dynamics (in light-cone time) of the tachyon on an unstable brane in the background 
of a dilaton linear along a null coordinate is a non-local reaction-diffusion type equation,
which admits a travelling front solution. We analyze the (in-)stability of this solution using
linearized perturbation theory. We find that the front solution obtained in singular perturbation
method is stable. However, these inhomogenous solutions (unlike the homogenous solution)
also have Lyapunov exponents corresponding to unstable modes around the (meta-)stable 
vacuum.}
\end{quote}
}

\bigskip\bigskip


\end{titlepage}
\maketitle\vfill \eject

\tableofcontents

\sectiono{Introduction}\label{sec:Introd}
It is interesting and important to study the dynamics of instabilities in string theory.  While this 
general question is too broad in its scope, the question of tachyonic instabilities in 
configurations of D-branes in string theory \cite{Sen:2004nf} is more specific and 
tractable\cite{Moeller:2002vx,Hellerman:2008wp,Joukovskaya:2008zv,%
Barnaby:2008pt,Beaujean:2009rb,Song:2010hc,Ghoshal:2011rs}. 
In particular, using light-cone coordinates and putting the open strings in a dilaton background,
which is linear along a null direction, the authors of Ref.\cite{Hellerman:2008wp} studied the 
homogeneous\footnote{Following standard terminology, by {\em homogeneous 
decay} we mean the dynamical evolution of the tachyon dependent on (light-cone) time only.} 
decay process in the effective field theory of the tachyon and extended this to a complete set 
of equations of motion of the open string field theory.
If we consider inhomogeneous decay in this framework in which the tachyon field depends
on the (light-cone) time and one other (spatial) coordinate along the brane, the equation of
motion of the tachyon turns out to resemble a reaction-diffusion type equation that was pioneered 
in Refs.\cite{Luther,Fisher, KPP} and appeared ubiquitously since. There are some additional
elements, however. Specifically, the non-linear reaction term of what we call the {\em Fisher 
equation for the tachyon on a decaying brane}, Eq.\refb{TachFishEqn}, involves a time delay 
and spatial averaging with a Gaussian kernel, hence it is non-local\cite{Ghoshal:2011rs}. Even
though non-locality in reaction-diffusion systems has been considered in the literature, in 
Mathematical Biology for instance (see \cite{BFish1,BFish2,Pattern} for example), the combination 
of delay and (the specific form of) non-local interactions that are inherent in open string field theory 
is quite characteristic. It also makes the resulting equations more interesting and difficult to analyze. 

As is the case for these type of equations, the Fisher equation for the tachyon also admits a 
{\em travelling front} solution. This front, which can be found using a singular perturbation 
analysis\cite{SPuri,MurrayMB,LDNath}, separates the brane from the (closed string) 
vacuum, while moving with a constant speed that is attained asymptotically. We have also
extended the traveling front to a solution of the equations of motion of open string field theory to 
the first non-trivial order\cite{Ghoshal:2013dla}. In terms of the boundary conformal field theory
on the worldsheet of the string, which provides the background for the open string field theory,
this corresponds to a deformation by a marginal operator which remains marginal when the 
first stringy corrections are included. The disc one-point functions of the closed string tachyon 
and graviton vertex operators, in the presence of this marginal deformation, were also studied 
in Ref.\cite{Ghoshal:2013dla}.

It is worth noting that the inhomogeneous decay described by the travelling front is closer to a
natural decay process. One would expect tachyon condensation to start, perhaps due to a 
fluctuation, in a small region of space. This nucleus, just like the condensation of a droplet in a
supercooled gas, would grow in size. In one dimension this would give rise to two fronts travelling
in opposite directions. In higher dimensions, the Laplacian $\nabla^2$ would appear in place of
$\partial_x^2$ in Eq.\refb{TachFishEqn} and the resulting equation is not quite a Fisher-type 
equation. However, for spherically symmetric decay, $\nabla_{d}^2 = \frac{\del^2}{\del r^2} + 
\frac{d-1}{r}\frac{\del}{\del r}$ is approximated by $\frac{\del^2}{\del r^2}$ for large $r$, leading
to a Fisher-type equation in the asymptotic limit. 

In this paper, we shall consider the stability of the traveling front. This analysis will be in the context
of the effective field theory of the tachyon. We shall study the behaviour of small fluctuations around
the front solution using linearized perturbation theory and argue that it is stable. We do, however,
find a potential instability around the stable vacuum, reminiscent of the oscillations in 
Ref.\cite{Moeller:2002vx}. This does not destabilize the front solution, obtained using a 
singular perturbation method starting with the solution corresponding to the homogeneous decay.

\section{Tachyon Fisher equation and the travelling front}
We recall that the dynamics of the open string modes are given by the cubic open string field theory.
In a given background, the string field can be expanded in terms of the states in the Hilbert space of
the underlying boundary conformal field theory on the worldsheet with coefficients that are the
`wavefunctions'. The leading contribution is the tachyon field $\phi(x^\mu)$ on the unstable brane.
This is a Klein-Gordon equation with negative mass-square augmented by {\em non-local} cubic 
self-interactions. The solutions of this equations are untamed oscillations\cite{Moeller:2002vx}
which may be attributed to the fact that the energy in the D-brane cannot be dissipated to the
closed string modes in the absence of any coupling between the open and closed string modes.

A simple and elegant approach to this problem that avoids the complexities of an open-closed 
string field theory was proposed in Ref.\cite{Hellerman:2008wp} and explored further by us
\cite{Ghoshal:2011rs,Ghoshal:2013dla}. The idea is to consider one of the light-cone coordinates
(say $x^+$) as time, and at the same time consider a dilaton background that in linear along 
the other light-cone direction $x^-$. This changes the essential character of the dynamical equations, 
while retaining the solvability of the underlying conformal field theory. In particular, the equation of 
motion of the tachyonic scalar field is
\begin{equation}
b\partial_t\phi - \partial_x^2\phi - m^2\phi
 + K^3 e^{-2\alpha b\partial_t + \alpha\partial_x^2}
\,\left( e^{\alpha\partial_x^2}\phi\right)^2 = 0,
\label{TachFishEqn}
\end{equation}
where $-m^2=-1$ is the mass-square of the tachyon, $b$ is the slope of the linear dilaton and 
$\alpha = \ln K = \ln \left( 3\sqrt{3}/4\right)$ is a number originating in the conformal maps that define 
the string field theory.  As mentioned above, $t \equiv x^+$ denotes light-cone time, and for simplicity, 
we have taken $\phi$ to depend only on one spatial coordinate $x$. This is a reaction-diffusion 
equation with time {\em delay} and spatial {\em non-locality}. We refer to it as the {\em Fisher equation 
for the tachyon on a decaying brane}. 

Like all equations of this type, of which there are innumerable examples in the literature, the above 
admits travelling front solutions. To see this, let us change variables to the comoving 
coordinate\footnote{This corresponds to a front moving to the left. The front moving to the right is
obviously also a solution.} and time
\[
\xi = x + vt,\qquad \tau = t,
\]
in terms of which the equation reads as follows:
\begin{equation}
b\frac{\del\phi}{\del\tau} + bv\frac{\del\phi}{\del\xi} - \frac{\partial^2\phi}{\del\xi^2} - \phi
 + K^3 e^{-2\alpha b\partial_\tau - 2\alpha bv\del_\xi + \alpha\partial_\xi^2}
\,\left( e^{\alpha\partial_\xi^2}\phi\right)^2 = 0.
\label{CoMovTachFishEqn}
\end{equation}
The travelling front does not have an explicit dependence on $t$ and is a function $\xi$ alone. 
Therefore it satisfies
\begin{equation}
bv\frac{\del\Phi_v}{\del\xi} - \frac{\del^2\Phi_v}{\del\xi^2} - \Phi_v + K^3 e^{-2\alpha bv\del_\xi 
+ \alpha\del_\xi^2}\,\left(e^{\alpha\del_\xi^2}\Phi_v\right)^2 = 0.
\label{leadingTachFish}
\end{equation}
The nonlocalities in the equations above can alternatively be written using the Gaussian kernel 
\begin{equation}
e^{\alpha\del_\xi^2}\, f(\xi) = \frac{1}{2\sqrt{\alpha\pi}}\displaystyle\int_{-\infty}^\infty d\xi'\, 
e^{-\frac{1}{4\alpha}\left(\xi' - \xi\right)^2}\, f(\xi') \equiv \frak{G}_\alpha[f(\xi)],
\label{Gauss}
\end{equation}
and the fact that $e^{-a\del_x} f(x) = f(x-a)$, (for which $e^{-a\del_x} \left(f(x)g(x)\right) 
= f(x-a)g(x-a)$ holds):
\begin{eqnarray*}
\left(b\frac{\del}{\del\tau} + bv\frac{\del}{\del\xi} - \frac{\partial^2}{\del\xi^2} - \right)\phi(\xi,\tau)
 + K^3\, \mathfrak{G}_\alpha\left[\left(\mathfrak{G}_\alpha\left[\phi(\xi-2\alpha bv,\tau-2\alpha b)
 \right]\right)^2\right] &=& 0,\nonumber\\
bv\frac{\del\Phi_v(\xi)}{\del\xi} - \frac{\del^2\Phi_v(\xi)}{\del\xi^2} - \Phi_v(\xi) + K^3
\mathfrak{G}_\alpha\left[\left(\mathfrak{G}_\alpha\left[\Phi_v(\xi-2\alpha bv)\right]\right)^2\right] &=& 0.
\end{eqnarray*}
The travelling front solution to these equations\cite{Ghoshal:2011rs} can be obtained in singular 
perturbation theory.

\subsection{Convergence and (in-)stability around the fixed points}
The differential equation above for inhomogeneous decay to leading order is of order two. However,
as in the case of the homogenous decay studied in Ref.\cite{Hellerman:2008wp}, it has two fixed 
points, and the travelling front interpolates between the unstable fixed point $\phi_U=0$ to the stable 
one at $\phi_S = K^{-3}\simeq 0.456$. It goes away from $\phi_U=0$ exponentially, the exponent 
being determined by the negative mass-square of the tachyon. Around $\phi_S$, however, due to 
the presence of the delay and non-locality, convergence is oscillatory. These can be deduced 
from a linearized perturbation analysis around the fixed points. 

\medskip

First, consider the unstable fixed point $\phi_U=0$. We can ignore the non-linear term in 
Eq.\refb{leadingTachFish} and substitute $\phi=e^{\mu\xi}$. This gives
\begin{equation}
\mu = \frac{1}{2}\left(bv \pm \sqrt{b^2v^2 - 4}\right)\nonumber
\end{equation}
which gives the minimum speed of the front as $bv=2$. The nonlocalities in the interaction term does 
not affect this behaviour, thus it is the same as in the standard Fisher equation. 

Indeed, this is true not only of the asymptotic speed, but also the way it is approached. Given 
a profile $\phi(x,0)$ at $\tau =0$, the solution to the equation linearized around $\phi_U=0$, 
namely $b \partial_{\tau} \phi = \partial^2_x  \phi + \phi$, is given by\footnote{A transformation 
$\phi\rightarrow e^{\tau/b}\phi$ brings it to the standard form of the diffusion equation.}
\begin{eqnarray}
\phi(x,\tau) &=& \displaystyle\int^\infty_{-\infty} \!\!dy \, \phi(y,0)\;
\frac{1}{\sqrt{-4 \pi \tau/b}}  e^{\frac{b}{4\tau} \left(-(x-y)^2 + \left(\frac{2\tau}{b}\right)^2\right)}
\nonumber\\
{}&\propto& \frac{1}{\sqrt{-4 \pi \tau/b}}\; e^{ -\frac{b}{4\tau}\xi^2 + \xi + {\cal O}(y) },
 \label{heatkernel}
\end{eqnarray}
where, we have rewritten the argument of the exponential in terms of the comoving coordinate with
the asymptotic velocity $\xi = x + \frac{2\tau}{b}$. (Note that the expression above is valid for 
$\tau\gtrsim -\infty$, near the unstable fixed point.) Now let $(\xi_{\phi_0},\tau)$ be the
coordinates at which the tachyon profile has reached a specific constant value $\phi_0$. Solving 
the equation above for $\xi_{\phi_0} (\tau)$, we obtain   
$\xi_{\phi_0} (\tau) \simeq \frac{1}{2}\ln \left(\frac{- \tau}{b}\right)$. 
Therefore, the asymptotic velocity is reached as $v(\tau) = v_{\mathrm{asym}} + 
\dot\xi_{\phi_0} \simeq  \displaystyle{\frac{2}{b}} - \displaystyle{\frac{1}{2 \tau}} + {\cal O} (\tau^{-2})$. 
While this is indeed the qualitative nature of the asymptotics, the coefficient of the $1/\tau$ term is 
not quite correct. This is because a derivative of the kernel of the diffusion equation also gives a 
solution, and in particular, taking the correction from the first derivative into account, we find
\begin{eqnarray}
\phi (x,\tau) &\propto & \left(x+\frac{2\tau}{b}\right)\; \exp\left[ -\frac{b}{4\tau}
\left(x+\frac{2\tau}{b}\right)^2 + \left(x + \frac{2\tau}{b} + \frac{3}{2} \ln \left(\frac{- \tau}{b}\right) 
\right) \right] \nonumber\\
v(\tau) &=& v_{\mathrm{asym}} + \dot\xi_{\phi_0} 
\: \simeq\:  \displaystyle{\frac{2}{b}} - \displaystyle{\frac{3}{2 \tau}} + \cdots.
\end{eqnarray}
We would like to reiterate that this analysis is exactly as in the case of the standard Fisher equation 
(see, for example, the review \cite{EbertS}) and is not affected by the non-local interactions. 

\medskip

On the other hand, the linearized equation for $\psi = \phi - \phi_S = \phi - K^{-3}$ around the stable 
fixed point differs from the standard case. The substitution $\psi=e^{\lambda\xi}$ in the linearized
equation leads to 
\begin{equation}
bv\lambda - \lambda^2 - 1 + 2 e^{2\alpha\lambda^2 - 2\alpha bv\lambda} = 0.
\label{Lyapunov}
\end{equation}
This is a transcendental equation which does not have any real solution, however, it admits an infinite 
number of complex solutions\footnote{The corresponding equation for the homogeneous case,
$b\lambda - 1 + e^{-2\alpha b\lambda} = 0$, is also a transcendental 
equation\cite{Hellerman:2008wp,Ghoshal:2011rs}. Its leading solutions are 
$-0.249613\, \pm\, i\, 1.90371$, however, $-3.91104 \,\pm\, i\,14.4748$, $-5.03573 \,\pm\, i\,26.7603$, 
$-5.73776 \,\pm\, i\,38.9404$, etc., which also satisfy the equation, are some of the non-leading
solutions.}, for example,  the leading behaviour is determined by 
\begin{equation}
\lambda = - 0.327933 \pm i\,0.716793
\label{uLyap}
\end{equation}
which differs slightly from the homogeneous case. (Some other solutions are
$-2.17775 \pm i\, 2.27752$, $-3.8092 \pm i\, 4.04854$, $-4.42422 \pm i\,4.70868$ etc.)
The exponent \refb{uLyap} is also not very different from the standard Fisher case, to which 
our equation reduces when $\alpha=0$: $(\lambda - 1)^2 = 2$, a solution of which is
$1- \sqrt{2} \simeq - 0.4142$. 

One should note, however, $1+\sqrt{2} \simeq +2.4142$ is also 
a solution of this quadratic equation --- the positive real part of $\lambda$ suggests that this 
corresponds to moving away from the stable fixed point $\phi_S$. However, in the standard 
analysis\cite{EbertS}, this positive exponent is eliminated by fixing the asymptotic conditions 
at $\xi\rightarrow\pm\infty$ determined by the front.

This potential instability is also present in the case of the tachyon. The equation for the exponent 
in Eq.\refb{Lyapunov} has a symmetry around $\lambda=1$ (for $bv=2$), and hence admits 
a solution $2.32793 \pm i\, 0.716793$ with a positive real part (and similarly for the other roots). 
The singular perturbation analysis that starts with the solution of the homogeneous equation as 
the seed, and thus fixes the asymptotic conditions at $\xi\rightarrow\pm\infty$, is not affected by 
this instability and yields a travelling front solution that converges. Nevertheless this instability 
could potentially cause the inhomogeneously decaying tachyon to oscillate around $\phi_S$ 
with increasing amplitude, the behaviour that was seen in the analysis of Ref.\cite{Moeller:2002vx}. 
In particular, as in Ref.\cite{Hellerman:2008wp}, one may attempt to solve Eq.\refb{leadingTachFish} 
by converting it into a recursion relation:
\begin{equation}
a_n = \frac{e^{\alpha(n^2 - 4n + 3)}}{(n-1)^2}\, \sum_{m=1}^{n-1} 
\left(a_m e^{\alpha m^2}\right)\,\left(a_{n-m} e^{\alpha (n-m)^2}\right)
\end{equation}
for $a_n$ in $\Phi_{bv=2}=\sum a_n e^{n\xi}$.  The coefficients increase rapidly, resulting in a
divergent series. 

\section{Perturbation of the non-local Fisher equation of the tachyon}
In this section, we shall analyze small fluctuations around the travelling front. To this end, let us 
separate the leading order front solution $\Phi_v$, that depends only on $\xi$, from the (small) 
perturbations around it
\[
\phi(\xi,\tau) = \Phi_v(\xi) + \eta(\xi,\tau), \quad | \eta | \ll | \Phi_v |.
\]
Thanks to Eq.\refb{leadingTachFish} satisfied by the leading order solution $\Phi_v$ (`classical solution'), 
the perturbations satisfy the linearized equation
\begin{equation}
b\frac{\del\eta}{\del\tau} + bv\frac{\del\eta}{\del\xi} - \frac{\partial^2\eta}{\del\xi^2} - \eta
 + 2 K^3 e^{-2\alpha b\partial_\tau - 2\alpha bv\del_\xi + \alpha\partial_\xi^2}
\,\left( e^{\alpha\partial_\xi^2}\Phi_v\right)\,\left( e^{\alpha\partial_\xi^2}\eta\right) = 0.
\label{PertEqn}
\end{equation}
where we have neglected terms of ${\cal O}(\eta^2)$. As expected, the translation zero-mode 
$\eta(\xi,\tau) = \del_\xi\Phi_v(\xi)$ is a solution to this.

Let us expand the perturbation $\eta(\tau,\xi)$ in terms of its (Fourier-Laplace) modes
\begin{equation}
\eta(\tau,\xi)  = \int_0^\infty dE \int_{- \infty}^{+ \infty} \frac{d p}{2 \pi}\, e^{-E \tau + i p \xi}\,  
\widetilde{\eta}_{E}(p),\nonumber
\end{equation}
but leave the front $\Phi_v$ as it is for the moment. If we make the plausible assumption that the operator 
$e^{-2 \alpha b \partial_{\tau} - 2 \alpha b  v \partial_\xi+ \alpha \partial_{\xi}^2 }$ in Eq.\refb{PertEqn} is invertible, 
we arrive at the equation: 
\begin{eqnarray}
\left( - b E +  p^2 -1 + i b v p  \right)
e^{ - 2 \alpha b E +  \alpha  p^2 + i 2 \alpha b v p  }\,  \widetilde\eta_E\left(p\right)
\!\!&=&\!\! 2 K^3  {\mathfrak G}_{\alpha}\! \left[ \Phi_v \right]  e^{- \alpha p^2 } \widetilde\eta_E\!\left(p\right), 
\nonumber\\
\left(\frac{1}{2}\frac{\partial}{\partial\alpha} {\cal L}_\alpha - 1\right) \widetilde\eta_E\left(p\right)
\!\!&=&\!\!  2 K^3 {\mathfrak G}_{\alpha}\! \left[ \Phi_v \right]  \widetilde\eta_E\!\left(p\right).
\label{eq:LPeta}
\end{eqnarray}
In the above, we have rewritten the equation in terms of a formal derivative  of the operator
\[ 
{\cal L}_\alpha = e^{ - 2 \alpha b E +  2\alpha  p^2 + i 2 \alpha b v p} \sim e^{2 \alpha b \partial_{\tau} +
2 \alpha b  v \partial_\xi - \alpha \partial_{\xi}^2 }
\]
with respect to $\alpha$ by an abuse of notation. (Recall that $\alpha = \ln\left(3\sqrt{3}/4\right)$ is a 
fixed number in OSFT.)

Following \cite{VolVoug}, let us consider the conditions for stability at asymptotic values of the front profile 
$\xi\rightarrow\pm\infty$. The Gaussian convolution ${\mathfrak G}_{\alpha} \left[ \Phi_v \right]$ of the front 
profile $\Phi_v$ softens the oscillations around the stable fixed point. 

As $\xi\rightarrow -\infty$, the tachyon profile $\Phi_v$ as well as its Gaussian transform 
${\mathfrak G}_{\alpha} \left[ \Phi \right]\rightarrow 0$. In this region, we have the operator equation 
$\partial_\alpha\left(\ln{\cal L}_\alpha\right) = 2$:
\begin{equation}
  b E =  (p^2 -1) + i b v p . \label{eq:EvalueAtLeft}
\end{equation}
This condition is exactly the same as in the case of the standard Fisher equation without any non-locality. 
This is not unexpected, as the behaviour of the two equations and their travelling front solutions are the 
same in this region. The travelling solution is said to be linearly stable if the perturbation decays 
exponentially in time, i.e., if $\mathrm{Re}(E) \geq 0$ (where $E =0$ corresponds to the translation 
zero-mode). Thus Eq.\refb{eq:EvalueAtLeft} may seem to indicate an instability at first sight because 
$\mathrm{Re}(bE) = p^2 - 1$, hence it is negative for $| p | < 1$. However, this is just the tachyonic 
instability at the maximum of the potential---the region $\xi\rightarrow -\infty$ still has the unstable 
D-brane.

Before we analyze the stability conditions in the asymptotic region $\xi\rightarrow\infty$ of the travelling 
front of the tachyon, let us review the situation for the usual Fisher equation, i.e., the case 
$\alpha \rightarrow 0$. Eq.\refb{eq:LPeta} reduces to a simple form:
\begin{eqnarray}
 b E  = \left(p^2 - 1 + 2 K^3 \Phi_v\right) + i b v p .  \label{eq:Fisheta}
\end{eqnarray}
Recall that at the true vacuum, $\Phi_v$ approaches the value $K^{-3}$. This means the solution is stable 
$\mathrm{Re}(bE) \approx p^2 + 1$ at the non-perturbative vacuum. (As mentioned above, in the region
$\xi\rightarrow -\infty$ corresponding to the perturbative vacuum $\Phi_v \approx 0$, the stability condition
is exactly the same with or without non-locality.) 

\begin{figure}[ht]
\begin{center}
\includegraphics[scale=0.50]{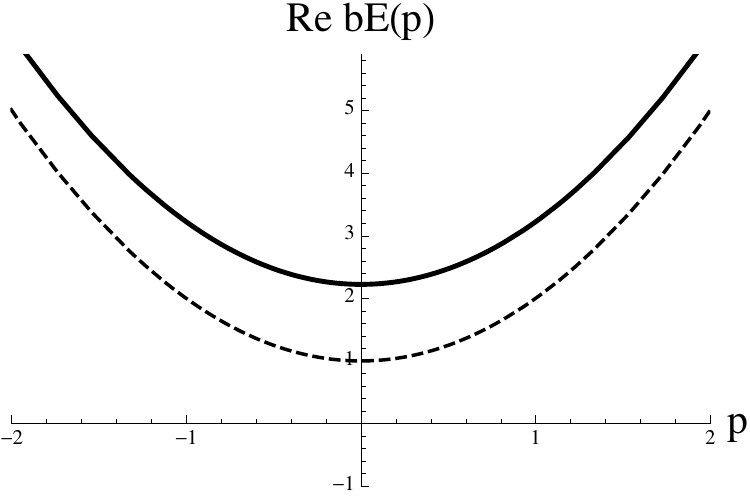}
\caption{{\small
Real parts of $b E({p})$, from Eqs.\refb{eq:Stringeta2} and \refb{eq:Fisheta}, for the travelling front of 
the tachyon Fisher equation and the standard Fisher equation around the non-perturbative vacuum 
(region where tachyon condensation has taken place) in dashed and solid lines, respectively. Both 
spectra are clearly non-negative, however, the tachyon front is even more stable than the travelling 
front of the standard Fisher equation.
}}
\label{fig:EEtaPlots}
\end{center}
\end{figure}

Getting back to general case with non-locality ($\alpha \neq 0$), the analytic form of the eigenvalue $E$ 
can be found by integrating the formal first order differential equation \refb{eq:LPeta} for the operator 
${\cal L}_\alpha$ with an integrating factor. By a straightforward integration of
\[
\frac{\partial}{\partial\alpha}\left(e^{-2\alpha} {\cal L}_\alpha \right) 
= 4 e^{-2\alpha} K^3 {\mathfrak G}_{\alpha}\! \left[ \Phi_v \right]  \widetilde\eta_E\!\left(p\right).
\]
we obtain
\begin{eqnarray}
b E = p^2 -1 + i b v p - \frac{1}{2 \alpha} \ln \left( 1 - 4 K^3 \int_{0}^{\alpha}{\mathfrak G}_{\alpha'}\!
\left[ \Phi_v \right] e^{-2 \alpha'} d \alpha' \right). \label{eq:Stringeta2}
\end{eqnarray}
This is valid for any value of $\xi$, and, in particular, the results for the region $\xi\rightarrow -\infty$ 
corresponding to the perturbative (unstable) extremum can be recovered. On the other hand, in the 
region $\xi\rightarrow\infty$, the front has settled to the stable (local) minimum  where 
${\mathfrak G}_{\alpha}\! \left[ \Phi_v \right] \approx K^{-3}$. Therefore, the argument of the logarithm 
can be approximated as $2 e^{-2 \alpha} -1$ which gives the real part of 
$\mathrm{Re}[b E] \approx p^2+2.22$. As a consequence, the travelling front of the tachyon is even 
more stable than the standard Fisher equation. The plots of $\mathrm{Re}\left(E\left(p\right)\right))$ 
for both cases are shown in Fig.\ref{fig:EEtaPlots}.

\subsection{Euclidean Schr\"odinger equation}
We can isolate the leading tachyonic instability around the perturbative vaccum from the effect of
fluctuations around the travelling front by the substitution
\[
\eta(\tau,\xi) = e^{bv\xi/2}\psi(\tau,\xi),
\]
which gets rid of the $\del_\xi\eta$ term in Eq.\refb{PertEqn} and brings the above to the form of an 
Euclidean Schr\"odinger equation. However, one should be careful due to subtelties that arise from the 
fact that $\psi$ does not belong to the Hilbert space of $L^2$-functions (because of the presence of the 
$\xi$-dependent prefactor). This is true of the standard Fisher case as well\cite{EbertS}. 

The equation satisfied by $\psi$ is
\begin{equation}
\begin{split}
b\frac{\partial\psi}{\del\tau} &= \frac{\partial^2\psi}{\del\xi^2} + \left(1 - \frac{1}{4}b^2v^2\right)\psi\\
&\qquad - 2 K^3 e^{-\frac{1}{2}bv\xi} \left[ e^{-2\alpha b v\partial_\xi + \alpha\partial_\xi^2} 
\,\left( e^{\alpha\partial_\xi^2}\Phi_v\right)\, \left(e^{\alpha\del_\xi^2 - 2\alpha b\del_\tau} 
e^{+\frac{1}{2} bv\xi} \psi\right)\right].
\end{split}
\label{towardScrodEqn}
\end{equation}
In order to simplify this further, we use the Campbell-Baker-Hausdorff formulas
to write
\begin{eqnarray*}
e^{\alpha\del_\xi^2}\, e^{bv\xi/2} &=& e^{\alpha b^2v^2/4}\, e^{bv\xi/2}\, 
e^{\alpha bv\del_\xi + \alpha\del_\xi^2}\\
e^{-2\alpha bv\del_\xi}\, e^{bv\xi/2}  &=& e^{- \alpha b^2v^2}\,e^{bv\xi/2}\, e^{-2\alpha bv\del_\xi}.
\end{eqnarray*}
This gives us the Euclidean Schr\"odinger equation for the perturbation function $\psi(\xi,\tau)$ as
\begin{eqnarray}
b\frac{\partial\psi}{\del\tau} &=& \frac{\partial^2\psi}{\del\xi^2} + \left(1 - \frac{1}{4}b^2v^2\right)\psi
\nonumber\\
&{}& \qquad - 2 K^3 e^{-\frac{1}{2}\alpha b^2v^2}\, \left[ e^{-\alpha b v\partial_\xi + \alpha\partial_\xi^2} 
\,\left( e^{\alpha\partial_\xi^2}\Phi_v\right)\, \left(e^{\alpha\del_\xi^2 + \alpha bv\del_\xi - 2\alpha b\del_\tau} 
\psi\right)\right]\nonumber\\
&=& \frac{\partial^2\psi(\xi,\tau)}{\del\xi^2} + \left(1 - \frac{1}{4}
b^2v^2\right)\psi(\xi,\tau)\nonumber\\
&{}& \qquad - 2 K^3 e^{-\frac{1}{2}\alpha b^2v^2}\, \frak{G}_\alpha\left[  \frak{G}_\alpha
[\Phi_v(\xi - \alpha bv)]\star \frak{G}_\alpha [\psi(\xi,\tau-2\alpha b)]\right].
\label{EScrodEqn}
\end{eqnarray}
(Notice that the argument of $\psi$ does not have a shift in $\xi$, though it has one in $\tau$.)
In the case of the standard Fisher equation without any non-locality, ($\alpha \rightarrow 0$) the above 
is a usual Schr\"{o}dinger equation:
\begin{eqnarray}
b \partial_{\tau} \psi = \partial_{\xi}^{2} \psi + \left( 1- \frac{1}{4} b^2 v^2 \right) \psi
- 2 K^3 \Phi_{v} \psi
\label{eq:FisherPert}
\end{eqnarray}
with the `potential' determined by the `classical' front solution $\Phi_v(\xi)$.

Let us point out some features of Eq.\refb{EScrodEqn}. The interaction with the `potential' $\Phi_v$ is 
non-local and in terms of a convolution product. Moreover, there is a delay in the argument of $\psi$ 
on the RHS of the above. Due to the delay, we do not get the conventional eigenvalue equation; rather 
writing $\psi(\xi,\tau) = e^{-E\tau}\,\Psi_E(\xi)$, the `time-independent' Schr\"odinger equation
\begin{equation}
\begin{split}
b E \Psi_E(\xi) &= - \frac{\partial^2\Psi_E(\xi)}{\del\xi^2} - \left(1 - \frac{1}{4}
b^2v^2\right)\Psi_E(\xi)\\
& \qquad + 2 e^{2\alpha b E}\,K^3 e^{-\frac{1}{2}\alpha b^2v^2}\, \frak{G}_\alpha\left[  \frak{G}_\alpha
[\Phi_v(\xi - \alpha bv)]\star \frak{G}_\alpha [\Psi_E(\xi)]\right].
\end{split}\label{EValuEqn}
\end{equation}
is a transcendental equation for $E$.
In order to show that the solution is stable, we need to prove that all the solutions to \refb{EValuEqn} have
$E\ge 0$. (Recall that $E=0$ is a solution that corresponds to translating the leading order solution.)

\begin{figure}[ht]
\begin{center}
\includegraphics[scale=0.50]{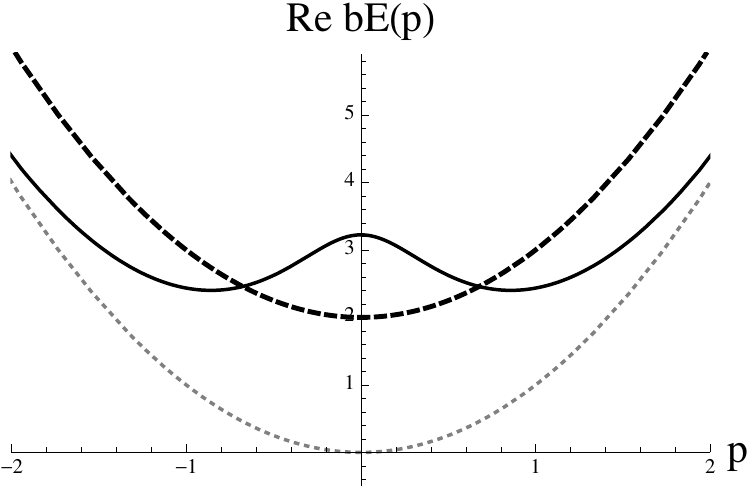}
\caption{{\small
Real parts of $b E({p})$, from Eqs.\refb{eq:Fishpsi} and \refb{eq:Stringpsi2}, for the travelling tachyon and 
standard Fisher cases around the perturbative (lower dotted curve --- exactly identical in both cases) and the 
non-perturbative vacua (upper curves: the solid one  for the tachyon and the dashed one for standard Fisher 
equation). For the travelling tachyon the minimum of the energy spectrum is at $p \approx \pm 0.86$.}}
\end{center}
\label{fig:EPsiPlots}
\end{figure}

In terms of the (Laplace-Fourier) modes
\begin{eqnarray}
\psi(\tau,\xi) &=&  \int_0^\infty dE \int_{- \infty}^{+ \infty} \frac{d p}{2 \pi} e^{-E \tau + i p \xi}\, \
\widetilde{\psi}_{E}(p),
\end{eqnarray}
Eq.\refb{eq:FisherPert} gives
\begin{eqnarray}
b E = p^2 +\left( \frac{1}{4} b^2 v^2 -1 \right) + 2 K^3 \Phi_v.
\label{eq:Fishpsi}
\end{eqnarray}
in the standard Fisher case.
It is obvious from Eq.\refb{eq:Fishpsi} that $E(p) \geq 0$ for all values of $p$. This is due to the factor of 
$\frac{1}{4} b^2 v^2$ on the right-hand side, and is expected from the form of the `potential' $\Phi_v$.
In the non-local case of $\alpha \neq 0$, we follow the same steps as in the analysis of $\eta$ to obtain:
\begin{equation}
b E = p^2 + \left( \frac{1}{4} b^2 v^2 -1 \right) 
- \frac{1}{2 \alpha} \ln \left[ 1 - 4 K^3 \int_{0}^{\alpha}\! d \alpha' \,{\mathfrak G}_{\alpha'}\!\left[ \Phi_v \right] 
e^{-2 \alpha' \left( 1 +i b v p \right)} \right]. 
\label{eq:Stringpsi}
\end{equation}
Similar to the conclusion for $\eta$, it turns out that at perturbative vacuum the spectrum is in the same form 
as Fisher case $b E = p^2 +\left( \frac{1}{4} b^2 v^2 -1 \right)$, and it is always non-negative. For 
non-perturbative vacuum, it may seem that non-negativity of $\mathrm{Re}[b E]$ is not guaranteed because
of the oscillations from $e^{i b v p}$.
However, at the non-perturbative vacuum ${\mathfrak G}_{\alpha}\!\left[ \Phi_v \right] \approx K^{-3}$, whence
Eq.\refb{eq:Stringpsi} reduces to
\begin{eqnarray}
b E = p^2  + \left( \frac{1}{4} b^2 v^2 -1 \right) - \frac{1}{2 \alpha}
\ln \left[ 1 - 2 \left(\frac{1-  e^{-2 \alpha \left( 1 +i b v p \right)}}{  1 +i b v p }\right)  \right]. 
\label{eq:Stringpsi2}
\end{eqnarray}
{}From the equation above, we find that the spectrum is symmetric, but not convex. The minimum of energy 
is not at zero: $\mathrm{Re}[b E] \approx 2.40$ at $p\approx\pm 0.86$. The oscillatory profile of $\Phi_v$ 
produces small potential wells, however, the positive contribution to the spectrum from the excitations can 
overcome the negative part from logarithmic term (of non-local effect). Numerical plots of real parts of spectrum 
$\mathrm{Re}[b E(p)]$ for both the standard Fisher and the travelling tachyon cases around the perturbative 
and non-perturbative vacua are shown in Fig.\ref{fig:EPsiPlots}

\begin{figure}[ht]
\begin{center}
\includegraphics[scale=0.55]{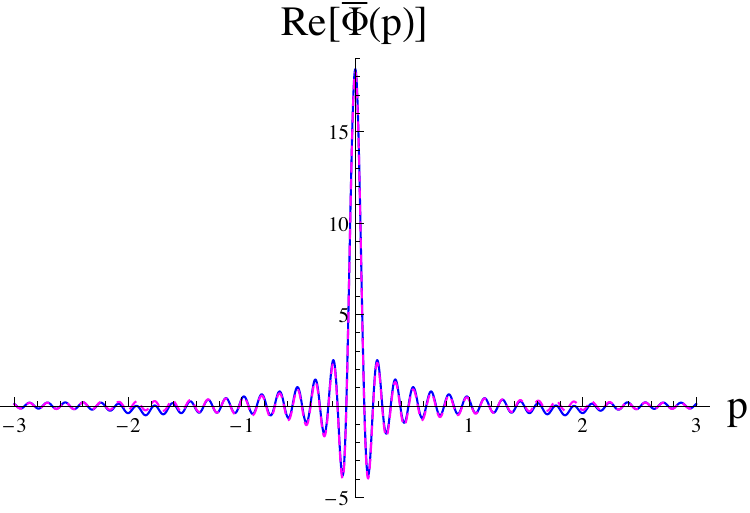}
\includegraphics[scale=0.55]{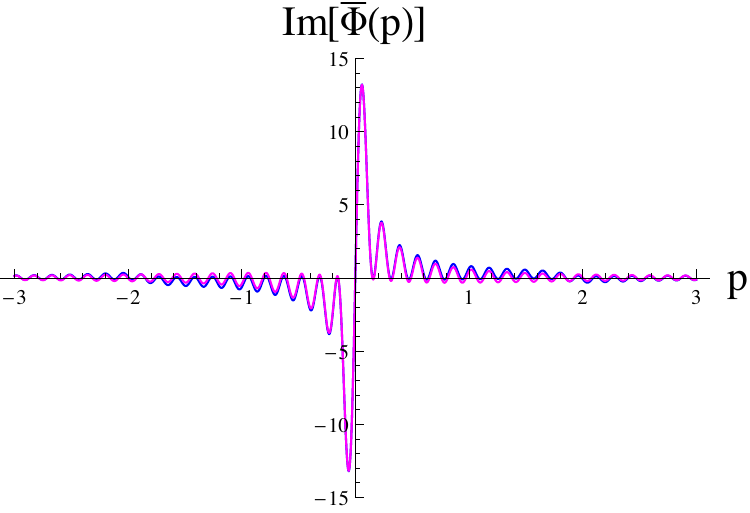}
\caption{{\small
Left: Real parts of Fourier transform of $\Phi_v^{(0)}$ for the tachyon Fisher and standard Fisher 
case in are shown in blue and magenta, respectively. The very small deviations between two plots 
occur in the region $1.4 \leq |p| \leq 2.2$.
Right: The corresponding imaginary parts. The very small deviations between the two plots are 
seen in the region $0.6 \leq |p| \leq 1.4$. }}
\end{center}
\label{fig:FourierPhi}
\end{figure}

In the above, we have taken the travelling front profile $\Phi_v$ to the leading order in the singular 
perturbation expansion (that is, we have worked with $\Phi_v^{(0)}$). However, it is straightforward 
to work with the profile including the effects at higher order. The qualitative behaviour is not expected 
to change. Plots for both the real and the imaginary parts of the Fourier transform of the tachyon front 
computed numerically are shown in Fig.\ref{fig:FourierPhi}. For this, we have put the system in a 
finite-size box (IR regulator). In spite of the oscillations around the stable vacuum, the difference from 
the standard Fisher case is small, and restricted to a finite region in momentum space.

\bigskip

One may also attempt to solve the non-local Schr\"{o}dinger equation by reducing it in to an integral 
eigenvalue problem. In terms of the modes $\widetilde{\Psi}_E(k)$:
\begin{equation}
\begin{split}
&\!\!\!\!\!\!\!\!\!\!\!\!  \left( b E  - k^2 + 1 - \frac{1}{4}b^2v^2 \right)\widetilde{\Psi}_E(k) \: = \\
& \qquad  2 K^3\, e^{2\alpha b E - 2\alpha k^2 - i\alpha bv k - \frac{1}{2}\alpha b^2v^2}\, 
\int\frac{d\ell}{2\pi}\, e^{2\alpha\ell(k-\ell) + i\alpha b\ell}\, 
\widetilde{\Phi}_v(k-\ell) \widetilde{\Psi}_E(\ell).
\end{split}\label{FTEValuEqn}
\end{equation}
Setting $\alpha=0$ in \refb{FTEValuEqn} recovers the standard Fisher equation (no delay or 
nonlocality) and its perturbation that satisfies
\begin{equation}
\left(b E - k^2 + 1 - \frac{1}{4}b^2v^2\right)\widetilde{\Psi}_E(k) \:=\:
2 K^3 \int\frac{d\ell}{2\pi}\, \tilde{\Phi}_v(k-\ell) \widetilde{\Psi}_E(\ell).
\label{FTEValuEqnFisher}
\end{equation}
The equations above are Fredholm integral equation of the second kind. 

As before, we may introduce
\[
{\cal U}_\alpha = \exp\left({-2\alpha \left(b E -  k^2 +1- \frac{1}{4} b^2v^2\right)}\right) \;\sim\;
\exp\left({2\alpha \left(b \partial_\tau + \partial_\xi^2 +1 - \frac{1}{4} b^2v^2\right)}\right)
\]
to write Eq.\refb{FTEValuEqn} compactly as
\begin{eqnarray*}
- \,\frac{\partial}{\partial\alpha}\,{\cal U}_\alpha\widetilde\Psi_E(k) &=&  4 K\, e^{ - i\alpha bv k}\, 
\int\frac{d\ell}{2\pi}\, e^{2\alpha\ell(k-\ell) + i\alpha b\ell}\, 
\widetilde{\Phi}_v(k-\ell) \widetilde{\Psi}_E(\ell)\\
\left(\mathbf{1} - {\cal U}_\alpha\right) \Psi_E(k) &=&  4 K\, \int\frac{d\ell}{2\pi}\, \left[\frac{e^{\alpha
\left(2\ell(k-\ell) + i b(\ell-vk)\right)} - 1}{2\ell(k-\ell) + ib \left(\ell - vk\right)}\right]\, 
\widetilde{\Phi}_v(k-\ell) \widetilde{\Psi}_E(\ell),
\end{eqnarray*}
where the last line is the result of integrating over the non-locality parameter from $0$ to $\alpha$, the
required value.

Before we close this section, since we have not come across it in the literature, it may not be entirely out 
of place to mention that the (Euclidean) Schr\"dingier equation for the perturbation of the standard Fisher 
equation, namely Eq.\refb{eq:FisherPert}, can be solved exactly to the lowest order in singular perturbation 
theory where $\Phi^{(0)}_v(\xi) = K^{-3}y(\xi) = K^{-3}/\left( 1+e^{-{\xi}/{b v}} \right)$. If we change variable 
from $\xi$ to $y$ and write 
\[
\psi(\tau,y(\xi)) = e^{-E\tau}y^\mu(1-y)^\nu F(y),
\]
then $F(y)$ satisfies a hypergeometric differential equation with parameters $a=(\mu+\nu)$, 
$b=(\mu+\nu+1)$ and $c=(2\mu+1)$, where $\mu^2 = -  b^2v^2\left(bE+1-\frac{b^2v^2}{4}\right)$ 
and $\nu^2 = -  b^2v^2\left(bE - 1-\frac{b^2v^2}{4}\right)$. 
We note in passing that the  the change of variable used above cannot be made in case of the 
tachyon Fisher equation, as the leading front is not monotonic due to non-local effects.

\section{Stability analysis in singular perturbation theory}
The travelling front solution of the tachyon Fisher equation \refb{leadingTachFish}, was solved by using a 
singular perturbation analysis\cite{Ghoshal:2011rs}, in which $\varepsilon = 1/(bv)^2$ was used as a small 
parameter. It is, therefore, natural to analyze the question of stability in this approach. In terms of the rescaled 
variable $\zeta = \sqrt{\varepsilon}\xi=bv\xi$ (and the derivative $bv\del_\xi = \del_\zeta$), used in the singular
perturbation theory, the equation for the perturbation \refb{PertEqn} takes the form
\begin{equation}
b\frac{\del\eta}{\del\tau} + \frac{\del\eta}{\del\zeta} - \varepsilon\frac{\partial^2\eta}{\del\zeta^2} - 
\eta + 2 K^3 e^{-2\alpha b\partial_\tau - 2\alpha\del_\zeta + \varepsilon\alpha\partial_\zeta^2}\,
\left( e^{\varepsilon\alpha\partial_\zeta^2}\Phi_v\right)\,\left( e^{\varepsilon\alpha\partial_\zeta^2}\eta
\right) = 0.\label{scaledPertEqn}
\end{equation}
Following the expansion of the leading order solution $\Phi_v(\xi) = 
\Phi_v^{(0)}(\xi) + \varepsilon \Phi_v^{(1)}(\xi) + \cdots$, we now expand the perturbation as well:
\[
\eta(\xi,\tau) = \eta^{(0)}(\xi,\tau) + \varepsilon\, \eta^{(1)}(\xi,\tau) + \varepsilon^2\, \eta^{(2)}(\xi,\tau) + \cdots
\]
Moreover, since the Gaussian kernel is identity for $\alpha=0$, it can be divided as\cite{Ghoshal:2011rs}
\begin{equation}
\frak{G}_{\varepsilon\alpha}[F(\zeta)] = F(\zeta) + \frak{dG}_{\varepsilon\alpha}[F(\zeta)],
\label{splitGauss}
\end{equation}
in which we treat $\frak{dg}\sim{\cal O}(\varepsilon)$. 

This leads to the following equations:
\begin{equation}
\begin{split}
{\cal O}(1):\quad& 
b\frac{\del\eta^{(0)}}{\del\tau} + \frac{\del\eta^{(0)}}{\del\zeta} - \eta^{(0)} + 2 K^3 
e^{-2\alpha b\partial_\tau - 2\alpha\del_\zeta}\,\left(\Phi_v^{(0)}\,\eta^{(0)}\right) = 0,\\
{\cal O}(\varepsilon):\quad& 
b\frac{\del\eta^{(1)}}{\del\tau} + \frac{\del\eta^{(1)}}{\del\zeta} - \eta^{(1)} + 2 K^3 
e^{-2\alpha b\partial_\tau - 2\alpha\del_\zeta}\,\left(\Phi_v^{(0)}\,\eta^{(1)}\right)\\
{}&\qquad = \;\frac{\del^2\eta^{(0)}}{\del\zeta^2} - 2 K^3 e^{-2\alpha b\partial_\tau - 2\alpha\del_\zeta}\,
\bigg(\frak{dg}_{\varepsilon\alpha}\left[\Phi_v^{(0)}\eta^{(0)}\right] + \Phi_v^{(0)} \left(\frak{dg}_{\varepsilon
\alpha}\left[\eta^{(0)}\right]\right)\\
{}&\qquad\qquad\qquad\qquad\qquad + \left( \left(\frak{dg}_{\varepsilon\alpha}\left[\Phi_v^{(0)}\right] + 
\Phi_v^{(1)}\right) \eta^{(0)} \right)\bigg)
\end{split}\label{OrderedPertEqn}
\end{equation}
plus equations for higher order terms. The first equation above for the leading term of the perturbation
is not a Schr\"{o}dinger-type equation being first order in time as well as the space derivatives. However, 
it is homogeneous, while the equations at second (and higher) order are inhomogeneous, with the sources 
determined from those at lower order.

Consider the Fourier transformed functions
\[
\Phi_v(\zeta)\; =\; \int \frac{dk}{2\pi}\, e^{ik\zeta}\,\widetilde{\Phi}_v(k),\qquad
\eta(\tau,\zeta)\; =\; \int dE \int \frac{dk}{2\pi}\, e^{-E\tau + ik\zeta}\,\widetilde{\eta}_E(k),
\]
which are valid at every order in perturbation. The equation at lowest order is
\begin{equation}
\left(-bE + ik - 1\right)\,e^{-2\alpha bE + 2i\alpha k}\,\tilde{\eta}_E^{(0)}(k) =
2K^3 \int\frac{d\ell}{2\pi}\,\tilde{\Phi}_v^{(0)}(k-\ell)\,\tilde{\eta}_E^{(0)}(\ell),
\label{FTPertZero} 
\end{equation}
which is again a Fredholm integral equation of second kind. The equations at higher order are 
Fredholm equation of first kind, consequently these may be solved by iterative technique. For the
standard Fisher equation ($\alpha=0$) one can once again change variable to 
$y = 1/\left(1+e^{-\zeta}\right)$, which leads to a simple quadrature
\begin{eqnarray*}
 \frac{d \eta^{(0)}}{d y} = \frac{1+ b E -2 y}{y(1-y)},
\end{eqnarray*}
integrating which we get $\eta^{(0)}(y) = K^{-3} y^{1+bE} (1-y)^{1-bE}$. 
We see that for $bE =0$, $\eta^{(0)}$ is a translation of the `classical' front $\Phi_{(v)}^{(0)}$
\begin{equation}
\eta^{(0)}\left(y(\zeta);E=0\right)\, =\, K^{-3}\, y (1-y)\, =\,  \frac{K^{-3}\,e^{-\zeta}}{(1+e^{-\zeta})^2} \,=\,
\frac{d}{d\zeta} \Phi_{v}^{(0)}(\zeta) \nonumber 
\end{equation}
as expected.

\section{Conclusions}
The dynamical equation of the tachyon on an unstable D-brane does not have a solution that interpolate 
between the extrema of the potential\cite{Moeller:2002vx}. However, in the background of a dilaton 
that is linear along a light-like coordinate $x^-$, the equation of motion (in light-cone time $x^+$) is 
first order. This admits an interpolating solution that has an oscillatory convergence to the (closed-string) 
vacuum\cite{Hellerman:2008wp}. This equation is actually a variant of a reaction-diffusion equation, 
which has nonlocal interactions, including a delay. Therefore, in the case of an inhomogeneous decay, 
there is a travelling front solution that moves with an asymptotic velocity converting regions of space 
from the unstable brane to the vacuum in its wake\cite{Ghoshal:2011rs}. In this paper, we have carried 
out a stability analysis of the front solution using linearized perturbation theory. The equations for the 
perturbation is a nonlocal Euclidean Schr\"{o}dingier equation, with the front profile acting as a potential. 
Thanks to the nonlocality, however, the potential and the `wavefunction' are in a convolution product. 
We find that the front solution found from a singular perturbation analysis is stable. We have also 
analyzed (linear) stability around the closed string vacuum. The Lyapunov exponents are determined 
by transcendental equations, which are different for the case of homogeneous and inhomogeneos
decay. For the latter, there are positive solutions that corresponds to (oscillatory) divergence. Even
though these modes do not destabilize the travelling front obtained in the singular perturbation
theory, their existence suggests that there could be space-time dependent solutions of the equation of
motion of the tachyon that exhibit untamed oscillation with increasing magnitude around the (closed-string) 
vacuum, similar to those of homogenous decay in usual time\cite{Moeller:2002vx}. The inclusion of the
higher string modes may change the dynamics --- we know that the tachyon perturbation corresponding 
to the front solution can be extended to the equations of string field theory to the next 
order\cite{Ghoshal:2013dla}.

\bigskip\bigskip

\noindent{\bf Acknowledgments:} It is a pleasure to thank Dushyant Kumar, Ravi Prakash, 
Sanjay Puri and Ram Ramaswamy. The work of DG was supported in part by SERC, DST 
(India) through the grant DST-SR/S2/HEP-043/2009, and PP is very grateful to Phra Jandee 
Jindatham and Watcharaporn Ladadok for inspiration and encouragement.

\bigskip\bigskip


\end{document}